\documentclass[epj]{webofc}
\usepackage[varg]{txfonts}   
\wocname{EPJ Web of Conferences}
\woctitle{INPC 2013}
\begin{document}
\title{Equilibrium and equilibration in a gluon plasma with improved matrix elements}
%
%

\author{Bin Zhang\inst{1}\fnsep\thanks{\email{bzhang@astate.edu}}
}

\institute{Department of Chemistry and Physics, Arkansas State University, P.O. Box 419, State University, AR 72467-0419, U.S.A.
}

\abstract{%
The hot and dense matter created in the early
stage of a relativistic heavy ion collision is composed mainly of
gluons. Radiative processes can play an important role for the
thermalization of such partonic systems. The simplest parton 
number changing processes are commonly described by the Gunion-Bertsch
formula. We show that the cross section from the exact matrix element
for the lowest order radiative process could be significantly smaller
than that based on the Gunion-Bertsch formula. In light of this, 
we discuss the role of radiative processes on the equilibrium and 
equilibration of a gluon plasma.
}
\maketitle
\section{Introduction}
\label{intro}
Relativistic heavy ion collisions have been used to produce and
study matter under extreme conditions similar to those existed
in the early Universe. Many spectacular properties
of hot and dense nuclear matter have been observed. With the
advancement of the hydrodynamical and related simulations, people
were able to recognize that the quark and gluon system quickly
thermalizes. The resultant Quark-Gluon Plasma, along with
some cold atomic systems, evolves with the lowest ever observed
shear viscosity to entropy density ratio \cite{Heinz:2013th}. 
Intensive research efforts have been directed toward understanding 
these observations \cite{Xu:2004mz,Asakawa:2006tc,Zhang:2008zzk,
Huovinen:2008te,El:2009vj,Denicol:2010xn,
Martinez:2010sc,Dusling:2012ig,Gelis:2013rba}. Some recent 
microscopic studies focused on contributions from particle number 
changing processes \cite{Xu:2007ns,Chen:2009sm,Zhang:2010fx,
Chen:2010xk,Zhang:2012vi,Fochler:2013epa,Huang:2013lia}. 
The Gunion-Bertsch formula \cite{Gunion:1981qs,Wang:1994fx}
is widely used to study the lowest order radiative process.
In the following, we will discuss our recent attempts at performing
simulations beyond the Gunion-Bertsch formula.

\section{Results and discussions}
\label{resul}

Our focus will be on the gluons as they form the dominant component
in the initial stage and they interact more strongly compared with
the quarks. Instead of using the Gunion-Bertsch formula, our 
calculations will start from the exact formula 
\cite{Gottschalk:1979wq,Berends:1981rb} for the lowest order 
radiative process. In this case, the matrix element modulus 
squared can be determined from the inner-products of particle 
four-momenta by
\begin{equation}
|M_{gg\rightarrow ggg}|^2=\frac{g_s^6N_c^3}{2(N_c^2-1)}
\frac{\sum (ij)^4\sum (ijklm)}{\prod (ij)}.
\end{equation}
Here, $g_s$ is the strong interaction coupling constant, and $N_c$
is the number of colors. The string $(ijklm)=(ij)(jk)(kl)(lm)(mi)$
where $(ij)$ is the inner-product of the four-momenta of
particles $i$ and $j$. The sums and product are over all distinct
permutations of particle labels, and the average over initial
and sum over final internal (spin and color) degrees of freedom have
been performed. The singularities in the denominator come from
propagators. We will regulate these propagators by the Debye
screening mass squared. Calculations can be done for a typical
gluon plasma like that produced in the early stage of a heavy
ion collision. When the strong interaction fine structure
constant $\alpha_s= g_s^2/(4\pi) =0.47$, the Debye screening
mass squared $\mu^2=10$ fm$^{-2}$, the two-particle
center-of-mass energy squared $s=4$ GeV$^2$, the calculated
two-to-two elastic scattering cross section
$\sigma_{22}=9\pi\alpha_s^2/(2\mu^2)=0.312$ fm$^2$, while the
two-to-three cross section $\sigma_{23}=0.0523$ fm$^2$.
The ratio $\sigma_{23}/\sigma_{22}= 16.8\%$ and is
significantly smaller than the ratios from the Gunion-Bertsch
formula based calculations which are about 
50\% \cite{Xu:2004mz}. When
$\alpha_s$ is changed to $0.3$, the corresponding
$\mu^2=6.38$ fm$^{-2}$. This leads to $\sigma_{22}=0.199$ fm$^2$
and $\sigma_{23}=0.0504$ fm$^2$. Again, the ratio is much
smaller than $50\%$. The Dalitz plot of the outgoing particles
shows that they are close to isotropically distributed. However,
the reaction integral and the outgoing particle distribution
can be very different from isotropic for the inverse process.

\begin{figure}[h]
\centering
\sidecaption
\includegraphics[width=10cm,bb=90 50 390 300,clip]{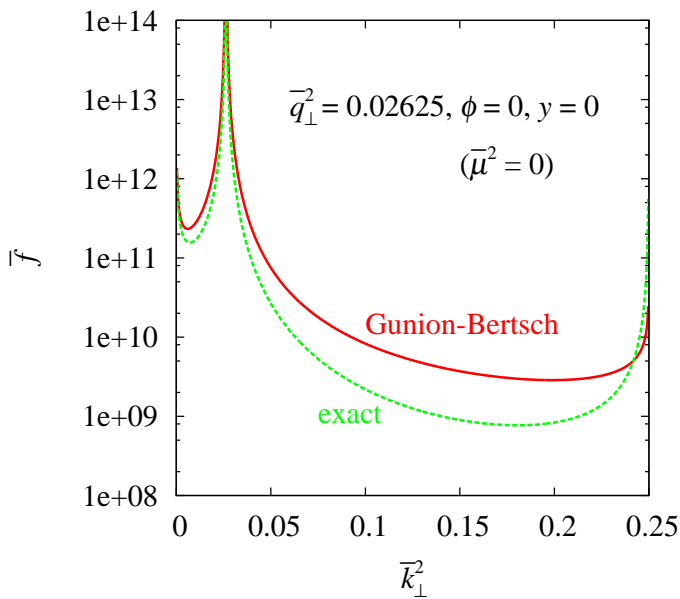}
\caption{Comparison of results from the Gunion-Bertsch formula and
the exact formula. The quantities are defined in the text.}
\label{fig-cmp1}       
\end{figure}

\vspace{-0.5cm}

It is interesting to see how the exact formula based calculations
compare with the Gunion-Bertsch formula based ones in different
phase space regions. The Gunion-Bertsch formula can be expressed
in terms of the transverse momentum transfer $\vec{q}_\perp$ and
the transverse momentum of the radiated gluon $\vec{k}_\perp$ as
\begin{equation}
|M_{gg\rightarrow ggg}^{GB}|^2=\frac{9g_s^4s^2}{2(q_\perp^2)^2}
\frac{12g_s^2q_\perp^2}{k_\perp^2(\vec{k}_\perp-\vec{q}_\perp)^2}.
\end{equation}
The singularities in the above formula can also be regulated
by the Debye screening mass squared. The comparison can be done
for $f(q_\perp,k_\perp,y,\phi)=
\sum_{y'_{1a},y'_{1b}}|M|^2/|\partial F/\partial y'_1|_{F=0}$.
Here $y$ is the rapidity of the radiated gluon, and $\phi$ is
the angle between $\vec{q}_\perp$ and $\vec{k}_\perp$. $y'_{1a}$
and $y'_{1b}$ are the two $y'_1$ values that solve $F=0$, where
$y'_1$ is the rapidity of momentum transfer $\vec{q}$, and $F=0$
is the mass shell condition for the third outgoing particle
(other than the radiated one or the momentum transferred one).
$f$ can be obtained by integrating out the energy-momentum
conserving delta function when calculating the cross section
or the rate per unit volume. Therefore, $f$ is proportional
to the differential cross section, and it reflects the outgoing
particle distribution. For any given set of $q_\perp$, $k_\perp$,
$y$, $\phi$ values, there are two sets of outgoing particles that
satisfy energy-momentum conservation. When $y=0$, they are
related by the mirror symmetry with respect to the transverse plane,
and the two configurations give the same exact matrix element
modulus squared. When $y\neq 0$, they generally give different
exact matrix elements. In this respect, $f$ allows a one-to-one
comparison between the Gunion-Bertsch formula and the exact
formula. Since $f$ is proportional to $\alpha_s^3/s^2$, it is
helpful to use the dimensionless quantity $\bar{f}=f/(\alpha_s^3/s^2)$.
Likewise, one can also rescale $q_\perp$, $k_\perp$, and $\mu$ by
$\sqrt{s}$ to obtain their dimensionless counterparts $\bar{q}_\perp$,
$\bar{k}_\perp$, and $\bar{\mu}$.
Fig.~\ref{fig-cmp1} compares $\bar{f}$ from the Gunion-Bertsch formula
and from the exact formula. The matrix elements were not regulated.
It is clear that $\bar{f}$ from the exact formula is smaller
than that from the Gunion-Bertsch formula anywhere but close
to the kinematic boundary at $\bar{k}_\perp^2=0.25$. The ratio 
increases from 1 at $\bar{k}_\perp^2=0$ 
to about 4 at $\bar{k}_\perp^2\sim 0.15$ and then drops down
to 0 as $\bar{k}_\perp^2\rightarrow 0.25$. 
This behavior cannot be accounted
for by the recently proposed correction factor for the Gunion-Bertsch
formula \cite{Fochler:2013epa}. At central rapidity ($y=0$), 
the ratio is not sensitive to $\bar{q}_\perp^2$, while the 
dependence on $\bar{q}_\perp^2$ can be large
at other places. A one-to-one comparison of the matrix element
modulus squared can be done at $y=0$. As expected, the Gunion-Bertsch
formula decreases continuously as $\bar{k}_\perp^2\rightarrow 0.25$,
missing the collinear singularity at $\bar{k}_\perp^2=0.25$.
Comparisons in other kinematic regions and with regulators can
also be done, and the Gunion-Bertsch formula and the exact formula
do not always agree.

\begin{figure}[h]
\begin{minipage}{65mm}
\includegraphics[width=65mm,bb=112 62 350 265,clip]{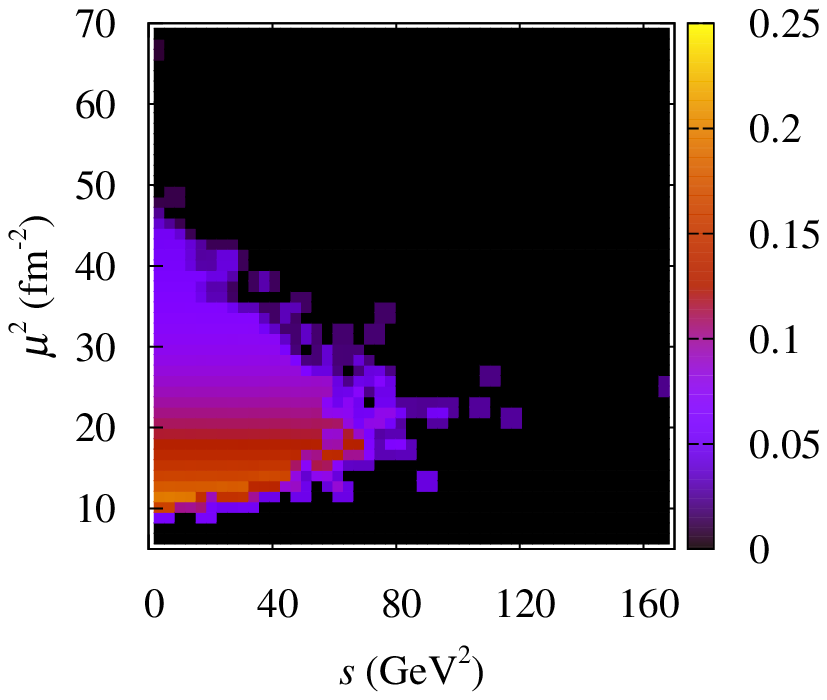}
\caption{The elastic cross section as a function of the Debye mass squared
and the center of mass energy squared. The colors of isolated points are
from averages of adjacent points and do not reflect the values at those
points.}
\label{fig-sig1}
\end{minipage}
\hspace{1pc}
\begin{minipage}{65mm}
\vspace{-1.2cm}
\includegraphics[width=65mm,bb=112 62 350 265,clip]{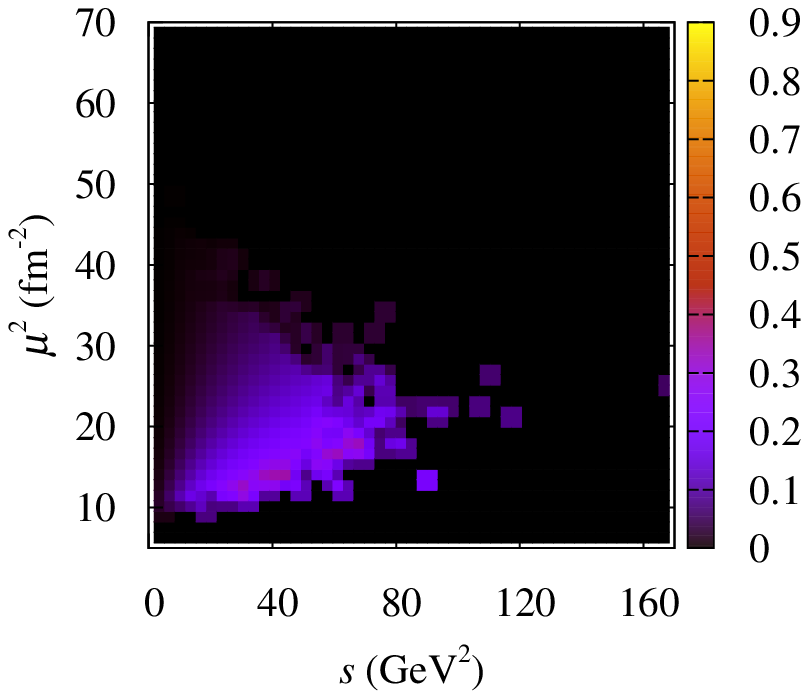}
\caption{Same as Fig.~\ref{fig-sig1} but for the 2 to 3 cross section.}
\label{fig-sig2}
\end{minipage}
\end{figure}

\vspace{-0.5cm}

In order to study the effect on the equilibration of a parton system,
the exact matrix element was implemented into our radiative transport
model. As a first step, we can look at the cross sections and rates
in equilibrium. In the following, we will show some preliminary results
and discussion their implications. The temperature will be set to
$T=0.524$ GeV, and the strong interaction fine structure constant
$\alpha_s=0.4$. It turns out that the 2 to 3 cross section averaged
over all 2 to 3 collisions is not that small relative to the
2 to 2 cross section averaged over all 2 to 2 collisions. The
ratio, $\langle\sigma_{23}\rangle_{23}/\langle\sigma_{22}\rangle_{22}\sim 
57\%$. However, if both the radiative and the elastic cross sections
are measured for every collision, the ratio $\langle\sigma_{23}\rangle/
\langle\sigma_{22}\rangle\sim 17\%$, significantly smaller than
the biased ratio of $57\%$. $\langle\sigma_{23}\rangle/
\langle\sigma_{22}\rangle$ is on the same order as that from the
typical cross section study in the beginning of this section.
Fig.~\ref{fig-sig2} clearly shows that the cutoff effect in the high
$\mu^2$ and low $s$ region for the radiative process relative to
the elastic process (Fig.~\ref{fig-sig1}). The ratio of the rates 
per unit volume, $w_{23}/w_{22}$, is even smaller 
than $\langle\sigma_{23}\rangle/\langle\sigma_{22}\rangle$. 
It returns a value around $12\%$ and shows the effect of particle 
distribution on the rates. There certainly can be alternative 
models for radiative processes. A recent study by the
Frankfurt group appears to agree qualitatively with our 
study \cite{Fochler:2013epa}.
If the radiative cross sections are smaller than previously 
expected from the Gunion-Bertsch formula,
chemical equilibration will also be slower. However, as particle
isotropization, kinetic equilibration, specific shear viscosity
depend also on the momentum transfer, detailed studies are 
necessary to find out how they behave.

\begin{acknowledgement}
We thank J. Mayfield for helpful discussions. This research was supported
by the U.S. National Science Foundation under Grant No. PHY-0970104.
\end{acknowledgement}

%
%
%

\end{document}